\documentclass[english,journal]{IEEEtran}
\usepackage[T1]{fontenc}
\usepackage[latin9]{inputenc}
\usepackage{xcolor}
\usepackage{pdfcolmk}
\usepackage{units}
\usepackage{amstext}
\usepackage{graphicx}
\PassOptionsToPackage{normalem}{ulem}
\usepackage{ulem}

\makeatletter

\providecolor{lyxadded}{rgb}{0,0,1}
\providecolor{lyxdeleted}{rgb}{1,0,0}

\DeclareRobustCommand{\lyxsout}[1]{\ifx\\#1\else\sout{#1}\fi}

\@ifundefined{date}{}{\date{}}
\usepackage{cite}
\usepackage[printonlyused]{acronym}
\acrodef{AWGN}{additive white Gaussian noise}
\acrodef{ASE}{amplified spontaneous emission}
\acrodef{QAM}{quadrature amplitude modulation}
\acrodef{PAM}{pulse amplitude modulation}
\acrodef{SE}{spectral efficiency}
\acrodef{SNR}{signal to noise ratio}
\acrodef{TX}{transmitter}
\acrodef{RX}{receiver}
\acrodef{BER}{bit error rate}
\acrodef{SER}{symbol error rate}
\acrodef{NFT}{nonlinear Fourier transform}
\acrodef{BNFT}{backward NFT}
\acrodef{FNFT}{forward NFT}
\acrodef{I-FNFT}{incremental FNFT}
\acrodef{DF-FNFT}{decision-feedback FNFT}
\acrodef{DF-BNFT}{decision-feedback BNFT}
\acrodef{NFDM}{nonlinear frequency-division multiplexing}
\acrodef{OFDM}{orthogonal frequency-division multiplexing}
\acrodef{NIS}{nonlinear inverse synthesis}
\acrodef{DAC}{digital-to-analog converter}
\acrodef{ADC}{analog-to-digital converter}
\acrodef{GVD}{group velocity dispersion}
\acrodef{SMF}{single mode fiber}
\acrodef{NCG}{Nystrom conjugate gradient}
\acrodef{B2B}{back-to-back}
\acrodef{EDC}{electronic dispersion compensation}
\acrodef{MAP}{maximum a posteriori probability}

\usepackage[linesnumbered,ruled,vlined]{algorithm2e}
\usepackage{algorithmic}

\makeatother

\usepackage{babel}
\begin{document}
\title{Sequence-Selection-Based Constellation Shaping for Nonlinear Channels}
\author{Stella Civelli\thanks{S.~Civelli is with the CNR-IEIIT, Pisa, Italy. S.~Civelli, E.~Forestieri,
and M.~Secondini are with the Telecommunications, Computer Engineering,
and Photonics Institute (TeCIP), Scuola Superiore Sant'Anna, Pisa,
Italy. E.~Forestieri, and M.~Secondini are also with the National
Laboratory of Photonic Networks, CNIT, Pisa, Italy. Email: stella.civelli@cnr.it.}, \IEEEmembership{Member, IEEE}, Enrico Forestieri, \IEEEmembership{Senior Member, IEEE},
and\\
 Marco Secondini, \IEEEmembership{Senior Member, IEEE}\thanks{This paper was presented in part at the Optical Communication Conference
(OFC), San Diego (CA), March 5-9, 2023. \cite{civelli2023invitedOFC,civelli2023practicalOFC}.}}
\maketitle
\begin{abstract}
Probabilistic shaping is, nowadays, a pragmatic and popular approach
to improve the performance of coherent optical fiber communication
systems. In the linear regime, the potential of probabilistic shaping
in terms of shaping gain and rate granularity is well known, and its
practical implementation has been mostly mastered. In the nonlinear
regime, the advantages offered by probabilistic shaping remain not
only valid, but might also increase thanks to the appealing opportunity
to use the same technique to mitigate nonlinear effects and obtain
an additional nonlinear shaping gain. Unfortunately, despite the recent
research efforts, the optimization of conventional shaping techniques,
such as probabilistic amplitude shaping (PAS), yields a relevant nonlinear
shaping gain only in particular scenarios of limited practical interest,
e.g., in the absence of carrier phase recovery. Recently, a more theoretical
approach, referred to as sequence selection, has been proposed to
understand the performance and limitation of nonlinear constellation
shaping. Sequence selection shapes the distribution of the transmitted
symbols by selecting or discarding the sequences generated by a certain
source according to a metric that measures their \emph{quality}. In
this manuscript, after a brief review of \emph{conventional} probabilistic
shaping, we use sequence selection to investigate through simulations
the potential, opportunities, and challenges offered by probabilistic
shaping for nonlinear channels. First, we show that ideal sequence
selection is able to provide up to $0.13$\,bit/s/Hz additional gain
with respect to PAS with an optimized blocklength. However, this additional
gain is obtained only if the selection metric accounts for the signs
of the symbols, ruling out the possibility of using one of the simple
recently proposed sign-independent metrics. We also show that, while
the signs must be known to compute the selection metric, there is
no need to shape them, since nearly the same gain can be obtained
by properly selecting the amplitudes (with a sign-dependent metric)
and leaving the signs uniform i.i.d. Furthermore, we show that the
selection depends in a non-critical way on the symbol rate and link
length: the sequences selected for a certain scenario still provide
a relevant gain if the link length or baud rate are modified (within
a reasonable range). Then, we analyze and compare several practical
implementations of sequence selection by taking into account interaction
with forward error correction (FEC), information loss due to selection,
and complexity. Overall, we conclude that the single block and the
multi block FEC-independent bit scrambling are the best options for
the practical implementation of sequence selection, with a gain up
to $0.08$\,bit/s/Hz. The main challenge and limitation to their
practical implementation remains the evaluation of the metric, whose
complexity is currently too high. Finally, we show that the nonlinear
shaping gain provided by sequence selection persists when carrier
phase recovery is included, in contrast to the nonlinear shaping gain
offered by optimizing the blocklength of conventional PAS techniques.
\end{abstract}


\section{Introduction\label{sec:Introduction}}

\IEEEPARstart{S}{}ince Shannon work in 1948, it is well known that
the capacity of an additive white Gaussian noise (AWGN) channel can
be theoretically achieved by mapping information to i.i.d. Gaussian
input symbols \cite{shannon48,shannon1949communication}. However,
for practical reasons, typical modulation schemes map information
to i.i.d. symbols drawn from a discrete modulation alphabet with uniform
distribution, such as the simple on-off keying employed by the first
generations of optical fiber systems, or more complex square $M$-ary
quadrature amplitude modulations (QAM) constellations employed by
more recent optical systems based on coherent detection, in which
$m=\log_{2}(M)$ bits are Gray-mapped to each QAM symbol.

The idea of approaching the Gaussian distribution with discrete constellations
has been discussed since the early 1990s \cite{calderbank1990nonequiprobable,kschischang1993optimal},
but only recently this concept gained considerable attention. Nowadays,
this strategy is known as constellation shaping and is a pragmatic
approach to improve the performance of communication systems. In particular,
probabilistic constellation shaping, or probabilistic shaping in short,
consists in mapping information bits on discrete square QAM constellations
with a desired probability distribution, chosen such as to approach
a Gaussian distribution \cite{bocherer2015bandwidth,buchali:jlt2016,cho2019probabilistic}.
In a nutshell, one decides to reduce the average energy per symbol
by using lower-energy symbols more frequently than higher energy ones.
The benefit obtained in terms of energy efficiency is partly paid
in terms of spectral efficiency: the redundancy of the source is increased
(its entropy decreased), and less than $m$ bits can be mapped to
each symbol. The best trade-off between energy and spectral efficiency
is obtained with a Maxwell--Boltzmann (MB) distribution, which, compared
to uniform QAM, allows to improve the performance of the system by
either obtaining the same information rate with a smaller signal-noise
ratio (SNR), or by improving the information rate for a given SNR,
closely approaching channel capacity. The way to map uniform i.i.d.
information bits to MB-distributed symbols, and to combine shaping
with forward error correction (FEC) has been deeply investigated in
the last years \cite{schulte2016CCDM,gultekin2020probabilistic,yoshida2019hierarchicalDM,civelli2020entropy,bocherer2015bandwidth,buchali:jlt2016},
and still is an active research topic nowadays.

Unfortunately, the optimal distribution for a \emph{nonlinear} fiber-optic
channel is still unknown and, most likely, not Gaussian \cite{secondini_JLT2017_scope}.
For this reason, in the recent years, a lot of effort has been devoted
to tailoring probabilistic shaping to the nonlinear fiber-optic channel,
trying to obtain an additional nonlinear gain on top of the linear
shaping gain \cite{geller:JLT2016,fehenberger:jlt2016,cho2021shaping}.
Indeed, it was recently shown that it is possible to further improve
the performance of a system beyond that achievable by merely tailoring
\emph{conventional} shaping techniques---developed for the linear
channel---to the nonlinear fiber channel. In particular, the \emph{sequence
selection} approach extends the conventional concept of probabilistic
shaping by automatically optimizing the source according to a proper
metric that measures the performance over a given channel. Even in
this case, the selection reduces the information rate of the source,
trading it for a higher robustness to nonlinear effects, with an overall
improvement of system performance \cite{civelli2021sequenceECOC,secondini2022new}.

In this paper, after a brief review of \emph{conventional} probabilistic
shaping in the linear and nonlinear regimes, we discuss the sequence
selection concept, providing interesting bounds for the performance
in the nonlinear regime and extending the analysis in \cite{civelli2023invitedOFC}.
We discuss the role of the signs of the transmitted symbols, which
are irrelevant in conventional probabilistic amplitude shaping (PAS)
but become fundamental for the nonlinear channel, and show that a
proper selection metric must account for the signs. Next, we propose
and discuss the implementation of probabilistic shaping techniques
for nonlinear channels based on sequence selection, briefly introduced
in \cite{civelli2023invitedOFC,civelli2023practicalOFC}, in particular
considering the interaction with forward error correction (FEC) and
carrier phase recovery (CPR), and considering different shaping and
concatenation strategies.

The manuscript is organized as follows. Section\,\ref{sec:System-description}
describes the system used for performance assessment in our simulations,
while Section\,\ref{sec:Probabilistic-constellation-shap} briefly
discusses probabilistic shaping for linear channels. Next, Section\,\ref{sec:Nonlinear-probabilistic-constell}
addresses probabilistic shaping for nonlinear channels, focusing on
(i) the performance of conventional shaping in the nonlinear regime,
(ii) the sequence selection approach, (iii) the practical implementation
of sequence selection, (iv) the performance of different sequence
selection approaches, and (v) the interaction with carrier phase recovery.
Finally, Section\,\ref{sec:conclusion} draws the conclusion.

\section{System Description\label{sec:System-description}}

The system setup is as follows \cite{civelli2023invitedOFC,civelli2023practicalOFC}.
We consider a dual polarization wavelength division multiplexing (WDM)
signal with $5\times\unit[46.5]{GBd}$ channels and $\unit[50]{GHz}$
spacing. The modulation format is dual polarization $64$-QAM with
probabilistic shaping and root-raised-cosine supporting pulse with
rolloff $0.05$. As a benchmark, and as a starting point (referred
to as unbiased source) for sequence selection, we consider probabilistic
amplitude shaping (PAS) with rate $R_{\text{4D}}=9.2$\,bit/4D (equivalent
to $R_{\text{DM}}=1.3$\,bit/amplitude), implemented with the serial
mapping for quadratures and polarizations \cite{civelli2023JLTNPN}.
The signal is sent into a link composed of $30\times100$\,km spans
of single mode fiber (SMF) with erbium-doped fiber amplifiers (EDFAs)
with noise figure $5$\,dB. At the receiver side, after demultiplexing,
the central channel undergoes electronic dispersion compensation,
matched filtering, symbol time sampling, and mean phase rotation compensation.
Finally, the achievable information rate (AIR) is evaluated as the
generalized mutual information (GMI), which represents the information
rate that can be achieved assuming ideal FEC and bit-wise mismatched
decoding under the bit metric decoding \cite{alvarado2018achievable,fehenberger2018multiset}.
The performance is  given in terms of spectral efficiency (SE) in
$\mathrm{bit/s/Hz}$, as $\mathrm{SE=46.5/50\cdot AIR}$. 

\section{Probabilistic Constellation Shaping\label{sec:Probabilistic-constellation-shap}}

Probabilistic constellation shaping---often simply referred to as
probabilistic shaping---denotes the way of mapping information bits
onto the symbols of a conventional $M$-QAM constellation such that
they turn out having a \emph{desired} prior probability. This is usually
done to approximate the capacity-achieving distribution and improve
system performance. In the case of an additive white Gaussian noise
(AWGN) channel with \emph{continuous} inputs, the capacity-achieving
distribution is Gaussian. On the other hand, when the inputs are constrained
on a given constellation of symbols $\mathcal{S}=\{\mathbf{s}_{1}$,$\mathbf{s}_{2}$,$\dots\}$,
an approximately optimal distribution is the Maxwell--Boltzmann (MB)
distribution\cite{kschischang1993optimal}\footnote{The MB distribution maximizes the entropy for a given constellation
and mean energy per symbol \cite{cover06}, and closely approach the
capacity for a given constellation and SNR.}
\begin{equation}
p(\mathbf{s})=\exp\left(-\lambda\Vert\mathbf{s}\Vert^{2}\right)/\sum_{\mathbf{s}_{\ell}\in\mathcal{S}}\exp\left(-\lambda\Vert\mathbf{s}_{\ell}\Vert^{2}\right)\label{eq:MB-distribution}
\end{equation}
where $\Vert\mathbf{\mathbf{s}}_{\ell}\Vert^{2}$ is the energy of
the symbol $\mathbf{\mathbf{s}}_{\ell}$, and $\lambda\geq0$ is a
parameter that characterizes the distribution and determines the trade-off
between bit rate and mean energy per symbol. 

In the past years, research has focused on the design of practical
approaches for the implementation of probabilistic shaping, leading
in particular to the development of the widely deployed probabilistic
amplitude shaping (PAS) technique \cite{bocherer2015bandwidth,buchali:jlt2016}.
The key idea of PAS is that of achieving probabilistic shaping by
concatenating a fixed-length-to-fixed-length distribution matcher
(DM) to shape the probability distribution of the amplitudes, and
a systematic binary encoder for forward error correction (FEC), whose
parity bits (possibly together with other information bits) are mapped
to the signs. The DM maps $k$ information bits (i.i.d. with uniform
probability) to $N_{\text{DM}}$ \emph{shaped} amplitudes from the
alphabet $\{1,3,\dots,2M_{\text{DM}}-1\}$ with a desired distribution
and should be invertible.\footnote{Sometime, it is useful to see the output of the DM in terms of bits,
simply considering a corresponding binary representation of the amplitudes
with $\lceil\log_{2}M_{\text{DM}}\rceil$ bits.} The reverse-concatentation scheme ensures that, at the receiver side,
FEC decoding can be performed before inverting the DM to demap the
information bits from the amplitude, so that the DM inversion is not
affected by error propagation and no iterations between demapper and
decoder are required. At the same time, the mapping of parity bits
on signs ensure that the distribution of the amplitudes is not altered
by the FEC, while the distribution of the signs remains uniform (as
required by (\ref{eq:MB-distribution})).

The rate of the DM $R_{\text{DM}}$ is the amount of bits that are
encoded on average on each amplitude, and equals $k/N_{\text{DM}}$.\footnote{The rate $R_{\text{DM}}=k/N_{\text{DM}}$ on the DM with an alphabet
of $M_{\text{DM}}$ positive amplitudes corresponds to the rate $R_{\text{4D}}=4(R_{\text{DM}}+1)$
in bit/4D on a dual polarization $M$-ary QAM constellation with $M=4M_{\text{DM}}^{2}$.} However, a practical DM with finite block length $N_{\text{DM}}$
is affected by a \emph{rate loss} $H-R_{\text{DM}}$\emph{ }(in bit/amplitude\footnote{The rate loss in bit/4D is simply the rate loss in bit/amplitude multiplied
by 4.})\emph{ }\cite{gultekin2018Sphereshaping}\emph{,} i.e., it works
at a rate that is slightly lower than the maximal rate $H$ (entropy)
at which information could be ideally mapped to i.i.d. amplitudes
with MB distribution and same energy per symbol.\footnote{Here, we adopt the definition given in \cite{gultekin2018Sphereshaping},
which, using the entropy of the MB distribution as a reference, is
directly related to the DM performance when applied to PAS over an
AWGN channel with an average power constraint. More general definitions,
which consider the entropy of an arbitrary target distribution or
of the empirical distribution at the output of the DM as a reference,
can also be found \cite{fehenberger2018multiset,yoshida2019hierarchicalDM,civelli2020entropy}.} For a given constellation size and block length $N_{\text{DM}}$,
the minimum rate loss (as defined above) is obtained with sphere shaping
(SpSh), which consists in using the $2^{k}$ sequences of length $N$
with minimum energy. Moreover, increasing the block length results
in a smaller rate loss. In practice, a DM with a reasonably low rate
loss requires a long block length (typically, hundreds of symbols),
making its implementation based on a single look-up-table (LUT) unfeasible.
Finding a good trade-off between rate loss and implementation complexity
is a key aspect in the design of DMs.Different techniques for the
implementation of DMs have been proposed in the past years, including
the enumerative sphere shaping (ESS) and its variations \cite{gultekin2017IEEE,gultekin2018Sphereshaping,gultekin2018approximate};
the constant composition distribution matcher (CCDM) and its variations
\cite{schulte2016CCDM,fehenberger2018multiset}; and the hierarchical
DM (HiDM), which, using multiple DM layers (each based, for instance,
on ESS, CCDM, or even LUTs), achieves a lower rate loss compared to
the equivalent-complexity single-layer DM \cite{yoshida2019hierarchicalDM,civelli2019hierarchicalOFC,civelli2020entropy}.
We refer to \cite{civelli2023JLTNPN} and references therein for more
details about DM implementations.

\begin{figure}
\includegraphics[width=1\columnwidth]{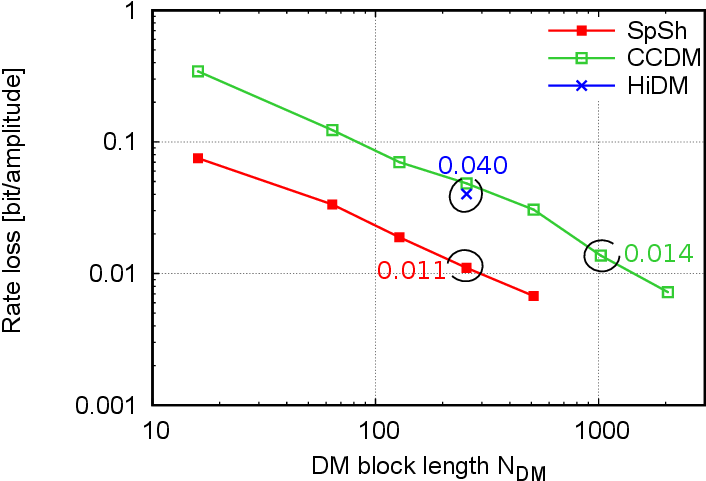}

\caption{\label{fig:rateloss} Rate loss versus block length.}
\end{figure}

As an example, the rate loss of SpSh and CCDM is shown in Fig.\ \ref{fig:rateloss}
as a function of the DM block length $N_{\text{DM}}$ for a rate of
$R_{\mathrm{DM}}\approx\unit[1.3]{bits/amplitude}$\footnote{The actual rate $R_{\text{DM}}$ is an approximation of $1.3$, since
it should be a ratio of two integer numbers. In particular, we select
$R_{\text{DM}}=\lfloor1.3N_{\text{DM}}\rceil/N_{\text{DM}}$.}. The figure also shows the rate loss of a LUT-based 6-layer HiDM
structure with a slightly larger rate $R_{\mathrm{DM}}=\unit[1.3164]{bits/amplitude}$
and characterized by the following LUT parameters per each layer:
output length $(8,2,2,2,2,2)$, number of input bits $(6,5,4,4,4,9)$,
output alphabet order $(4,64,\dots,64)$. The memory required to
save and store the LUTs of this HiDM is only $133$\,kbit, while
the number of operations required for the encoding and decoding is
negligible. In the following we will consider the three different
DMs highlighted in Fig.\,\ref{fig:rateloss}: the SpSh with block
length $256$, the CCDM with block length $1024$, and the already
mentioned HiDM. The CCDM and SpSh have approximately the same rate
loss but different block length (i.e., similar linear performance
but different nonlinear performance, as shown in the next section)
and a non-negligible complexity. On the other hand, HiDM has the same
blocklength as SpSh but higher rate loss and almost negligible complexity.
For a comparison of SE in the linear regime, please see the comments
about Fig.\,\ref{fig:performanceDM} in the next Section.

\section{Probabilistic Constellation Shaping for Nonlinear Channels\label{sec:Nonlinear-probabilistic-constell}}

Given the success of PAS in the linear regime, research has recently
been directed towards the possibility of using PAS also for nonlinearity
mitigation and compensation. In this context, the additional gain
(on top of the linear shaping gain) provided by PAS in the nonlinear
regime is often referred to as nonlinear shaping gain. In this Section,
we first discuss the performance of conventional DMs---i.e., DMs
that are designed to implement PAS and optimize its performance in
linear regime---in the nonlinear regime and briefly mention some
metrics proposed for performance prediction. Next, we describe the
sequence selection technique and compute the achievable performance
with different flavors of sequence selection. Then, we propose different
implementations of sequence selection, highlighting the main advantages
and disadvantages and showing their performance. Finally, we analyze
the interaction of sequence selection and carrier phase recovery.

\subsection{Nonlinear Performance of Conventional DM}

\begin{figure}
\includegraphics[width=1\columnwidth]{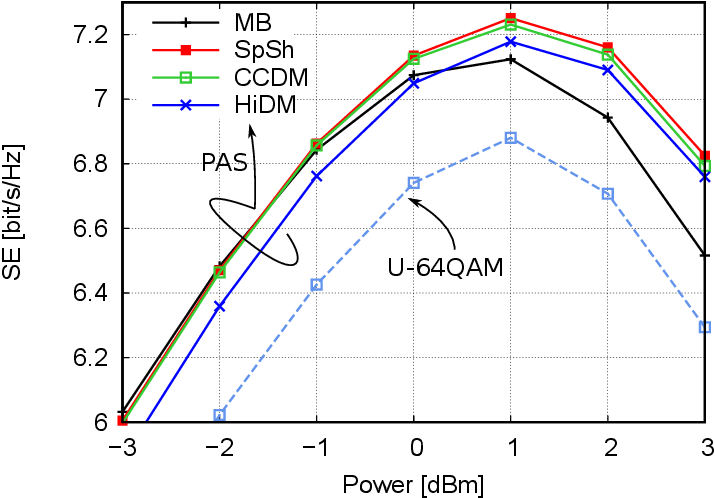}

\caption{\label{fig:performanceDM} SE versus power with conventional DMs.}
\end{figure}

The performance of ideal PAS (i.e., with i.i.d. MB-distributed symbols)
in the nonlinear regime has been investigated in \cite{fehenberger:jlt2016},
showing that even if the MB distribution is no longer optimal in this
case, a full numerical optimization of the target distribution in
a 1D (independently per each quadrature component) or 2D (jointly
for the in-phase and quadrature components) space does not yield any
performance improvement. On the other hand, it has been shown that,
in the nonlinear regime, non-ideal PAS based on short-block-length
DMs (e.g., SpSh and CCDM) provides an additional nonlinear shaping
gain compared to ideal PAS \cite{geller:JLT2016,amari2019introducing,fehenberger2020mitigating}.
Indeed, while increasing the block length of these DMs reduces the
rate loss and improves the linear performance,\footnote{For $N\rightarrow\infty$, both SpSh and CCDM converge to an ideal
source of i.i.d. MB-distributed symbols, and their rate loss vanishes.} using a shorter block length reduces the amplitude (energy) fluctuations
in the transmitted signal and induces less nonlinear interference
\cite{borujeny2023constant}. The optimal block length and performance
are obtained as a trade-off between these two effects and depend not
only on the considered scenario, but also on the specific DM implementation.
SpSh has better linear performance with respect to CCDM, band-limited
ESS \cite{gultekin2022mitigating}, Kurtosis limited ESS \cite{gultekin2021kurtosis},
and single-shell mapping \cite{civelli2023JLTNPN}, but has more energy
variations and a larger Kurtosis, which are responsible to generate
part of the nonlinear interference. Despite this, it was shown that
SpSh is able to achieve the largest nonlinear shaping gain, because
its lower rate loss allows to improve the linear performance and to
use a shorter block length \cite{civelli2023JLTNPN}. In the past
years, some metrics have been introduced to understand and predict
the performance of PAS in the nonlinear regime. With small differences,
the energy dispersion index (EDI) \cite{wu2021temporal}, the exponentially
weighted EDI \cite{wu2021EEDI}, the Kurtosis \cite{cho2021shaping},
and the lowpass-filtered symbol-amplitude sequence (LSAS) \cite{askari2023probabilisticJLT}
have been shown to predict the nonlinear interference in the system
and the received signal-noise ratio (SNR). All these metrics account
only for the energy of the symbols, i.e., they depend only on their
amplitudes and not on their signs.

The nonlinear behavior of SpSh, CCDM, and HiDM is shown in Fig.\,\ref{fig:performanceDM}
and compared with i.i.d. MB symbols and uniform 64QAM (U-64QAM). The
figure shows that in the linear regime, the best performance is obtained
with MB symbols, closely followed by SpSh and CCDM, while the HiDM
has the worst performance. This strictly follows the behaviour in
Fig.\,\ref{fig:rateloss}: at $\unit[-3]{dBm}$, the SE difference
between HiDM and MB is about 0.15~bit/s/Hz, approximately equal to
the HiDM rate loss of 0.16~bit/4D multiplied by $46.5/50$ to translate
it to bit/s/Hz. For SpSh and CCDM, the rate losses are significantly
lower and result in hardly noticeable SE differences in Fig.~\ref{fig:performanceDM}.
Conversely, in the nonlinear regime, SpSh achieves the best performance,
and i.i.d. MB symbols the worst. The HiDM approaches SpSh and CCDM
for increasing power. As expected, U-64QAM has the worst performance,
besides also having a higher rate ($6$ bit/2D rather than $4.6$
bit/2D).

However, the performance improvement obtained by employing short-block-length
DMs becomes nearly irrelevant when carrier phase recovery (CPR) is
included in the system, as in all practical systems. Indeed, we have
shown that also CPR mitigates some nonlinear interference---in fact,
the same nonlinear interference that is avoided by reducing the PAS
blocklength---improving the performance of all DMs and hiding the
nonlinear shaping gain provided by short block length PAS in many
practical scenarios \cite{civelli2023JLTNPN}. This behaviour is predicted
by the nonlinear phase noise (NPN) metric, which, differently from
EDI, Kurtosis, and LSAS, allows to account also for the impact of
CPR \cite{civelli2023JLTNPN}.

\subsection{Sequence Selection\label{subsec:Sequence-selection}}

\begin{figure}
\centering\includegraphics[width=1\columnwidth]{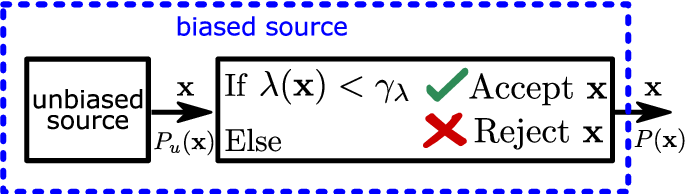}

\caption{\label{fig:seqsel}Sequence selection.}
\end{figure}

Sequence selection was proposed for the first time in \cite{civelli2021sequenceECOC}
as a novel technique to understand the potential and limitations of
shaping for nonlinear channels. Combined with an improved detection
metric, sequence selection has been shown to provide a new lower bound
to the capacity of the nonlinear optical fiber channel \cite{secondini2022new}.
In a nutshell, sequence selection uses only \emph{good} sequences
for information transmission, rejecting all others.

Sequence selection is implemented as a rejection sampling algorithm,
as in Fig.\,\ref{fig:seqsel}. An unbiased source generates sequences
$\mathbf{x}$ of $n$ 4D symbols, with unbiased probability $p_{u}(\mathbf{x})$,
e.g., with uniform distribution or probabilistic shaping. Given $\mathbf{x}$,
a selection metric (or cost function) $\lambda(\mathbf{x})$ is evaluated,
and the sequence is accepted and used for transmission if the metric
is below a certain threshold $\gamma_{\lambda}$. Note that in this
case both amplitudes and signs are shaped by the selection. The acceptance
rate $\eta$ is estimated as $\eta=N_{a}/N_{p}$, with $N_{a}$ and
$N_{p}$ being the number of accepted and proposed sequences. The
use of the biased source in Fig.\,\ref{fig:seqsel} induces a loss
in information, because a smaller set of all possible sequences is
used. This loss is accounted for in the achievable information rate
(AIR) as 
\begin{equation}
\text{AIR}=\text{AIR}_{u}-\frac{1}{n}\log_{2}\frac{N_{p}}{N_{a}}\,\,\,[\text{bits}/4\text{D}]\label{eq:air}
\end{equation}
where $\text{AIR}_{u}$ is the AIR according to the unbiased source
$p_{u}(\mathbf{x})$ and the second term is the loss due to selection
\cite[eq. (8)]{secondini2022new}. We refer to the procedure described
above as ``ideal'' sequence selection, meaning that it is not conceived
for a practical implementation (in contrast with the more practical
sequence selection schemes described in Section~\ref{subsec:Practical-implementation-seqsel})
but, rather, just as a way to estimate the AIR. The effectiveness
of sequence selection depends on the selection metric $\lambda(\mathbf{x})$,
as well as on the unbiased distribution $p_{u}(\mathbf{x})$, on the
sequence length $n$, and on the acceptance rate $\eta$. In the ideal
sequence selection case, to assess the potential of shaping for the
nonlinear channels and understand how much can be achieved, we use
the average NLI metric \cite[eq. (12)]{secondini2022new}, computed
according to the algorithm described in \cite[Sec. V-C]{secondini2022new}.
Though generally suboptimal in terms of AIR maximization,\footnote{The derivation of optimal and suboptimal metrics for the maximization
of (\ref{eq:air}) is discussed in \cite[Sec. III-D]{secondini2022new}.
For instance, it is shown that in the linear regime, when the NLI
becomes irrelevant and the AIR maximization problem is usually complemented
with an average-power constraint, the sequence energy would be a more
meaningful metric.} the metric provides an accurate estimate of the NLI in a single-channel
scenario, including the average impact of inter-block NLI due to adjacent
sequences, and is a good selection metric in the nonlinear regime
when an AWGN decoding metric is employed at the receiver \cite[Sec. III-D]{secondini2022new}.
Within the validity limits of a first-order regular perturbation,
the NLI metric scales cubically with the optical power. Thus, there
is no need to compute the metric at the actually employed launch power,
since the ranking induced by the metric on a set of sequences is approximately
independent of it \cite{secondini2022new}.

Ideal sequence selection provides an AIR, hence a lower bound to the
capacity of the channel under consideration. Ideally, with infinite
computation capabilities, sequence selection would allow to capture
the ultimate potential of shaping in the nonlinear regime, when the
acceptance rate tends to zero, the sequence length to infinity, and
the metric is perfect. Conversely, ideal sequence selection serves
as an upper bound to the performance of its practical implementations---discussed
in the next sections---with the same characteristics (e.g., acceptance
rate, sequence length), and will be used to infer conclusions about
probabilistic shaping in the nonlinear regime.

To understand how the potential of sequence selection depends on the
adopted metric, we consider three different cases. In the first case,
which we refer to as ideal sequence selection with \emph{shaped} signs,
we assume that both the amplitudes and signs are shaped according
to the average NLI metric $\lambda(\mathbf{x})$ described above.
In the second case, which we refer to as ideal sequence selection
with\emph{ unshaped} and \emph{unknown }signs, we assume that only
the amplitudes are shaped (as in PAS), and that the metric $\lambda(\mathbf{x})$
is independent of the signs of the symbols in $\mathbf{x}$. This
is the case when a simple metric is adopted (e.g., EDI, NPN, LSAS,
or Kurtosis) and/or when the signs are determined by the FEC after
the selection procedure, as in the reverse concatenation scheme of
PAS. As an ideal metric for this case we adopt the average NLI metric
described above, further averaging over the possible signs of the
symbols in $\mathbf{x}$. Finally, in the third case, we consider
the intermediate scenario of ideal sequence selection with \emph{unshaped}
and \emph{known} signs, where the signs of the symbols are known and
used to compute the metric $\lambda(\mathbf{x})$ as in the first
case, but they are fixed a priori and cannot be modified by the selection
procedure.

\begin{figure}
\centering \hfill{}(a)\hfill{}

\includegraphics[width=1\columnwidth]{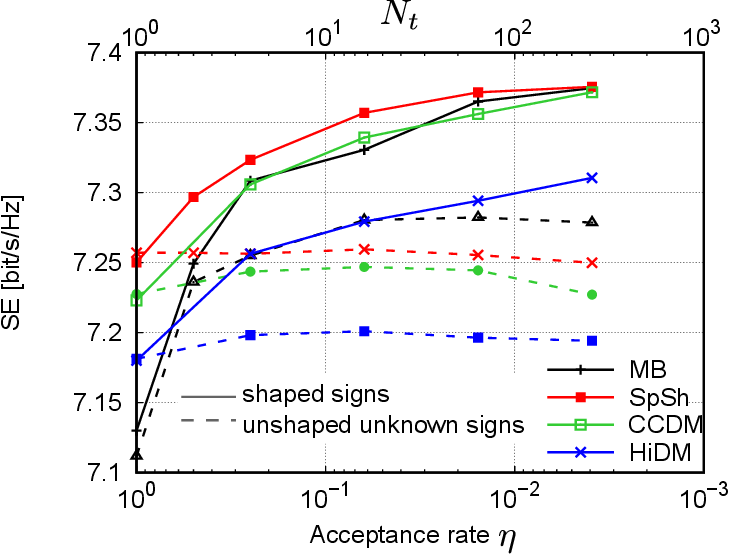}

\centering \hfill{}(b)\hfill{}

\includegraphics[width=1\columnwidth]{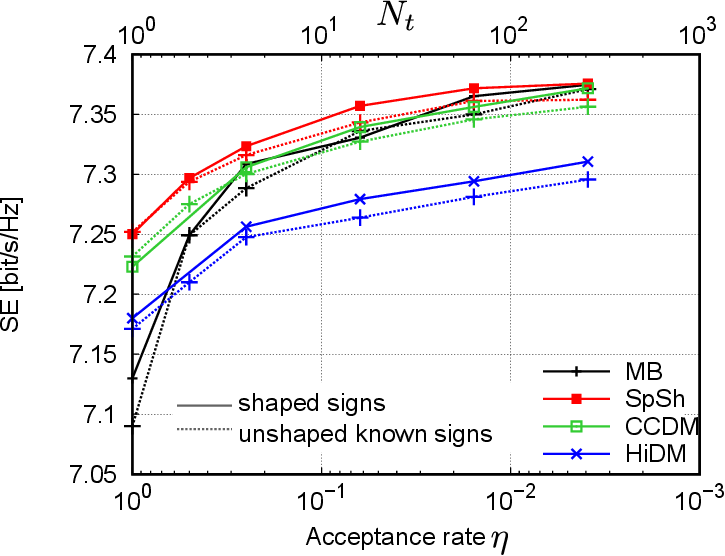}

\caption{\label{fig:results_seqsel} Optimal SE versus acceptance rate $\eta$
with ideal sequence selection with (a) shaped signs (solid) and unshaped
unknown signs (dashed), and (b) shaped signs (solid) and unshaped
known signs (dotted).}
\end{figure}
Figures\,\ref{fig:results_seqsel}(a)-(b) show the performance of
the system with different sequence selection strategies as a function
of the acceptance rate $\eta$, with $N_{p}=36864$ and at the optimal
power $P=\unit[1]{dBm}$.\footnote{Note that $P=\unit[1]{dBm}$ is not the exact optimal power, but only
an approximation. In the following, we will always consider the approximately
optimal power, meaning that we verified that it was optimal when taken
with $1$\,dBm steps.} Furthermore, in order to ease comparison with the results in the
next sections, the quantity $N_{t}=N_{p}/N_{a}$ is reported on the
top axis. As the unbiased source, we consider either i.i.d. MB-distributed
symbols, which yield the best performance in the linear regime with
no rate loss, or the three practical DMs whose rate loss is highlighted
in Fig.~\ref{fig:rateloss}. Obviously, when the acceptance rate
is $100$\%, i.e., sequence selection is not applied, SpSh, HiDM,
CCDM, and MB have the same performance shown in Fig.\,\ref{fig:performanceDM}
(at optimal launch power), regardless of the shaping strategy adopted
on the signs (but for minor statistical fluctuations). When sequence
selection is performed on both amplitudes and signs, the performance
of all PAS techniques improves, as shown by the solid line curves
in Figs.\,\ref{fig:results_seqsel}(a)-(b). Interestingly, even though
the four PAS strategies have different performance without sequence
selection, three of them (SpSh, CCDM, and MB) achieve nearly the same
peak performance $\mathrm{SE=7.38\,bit/s/Hz}$ when the acceptance
rate is reduced to the lowest tested value $\eta=3.9\cdot10^{-3}$,
with gains of 0.13~bit/s/Hz, 0.16~bit/s/Hz, and 0.25~bit/s/Hz,
compared to their respective values without sequence selection. This
is explained by the fact that the initial gap between the three techniques
depends on their different blocklength (256, 1024, and virtually infinite,
respectively), where shorter values mean less intensity fluctuations
and, consequently, less NLI. In this case, the sequence selection
procedure automatically controls also the intensity fluctuations of
the selected sequences, closing the performance gap between them.
On the other hand, HiDM has the same blocklength as SpSh but higher
rate loss (see Fig.~\ref{fig:rateloss}). This means that its intensity
fluctuations are already quite limited (as in SpSh), so that it gets
from sequence selection nearly the same gain as SpSh, while the initial
performance gap, which in this case is due to the rate loss, cannot
be recovered by sequence selection.

Clearly, the signs of the symbols in $\mathbf{x}$ are irrelevant
to determine the intensity fluctuations that cause NLI. This means
that the portion of the gain of sequence selection that can be ascribed
to the reduction of intensity fluctuations should be equally achieved
by shaping only the amplitudes with a sign-independent metric. This
is confirmed by the dashed-line curves in Fig.~\ref{fig:results_seqsel}(a),
which show that sequence selection with unshaped unknown signs provides
a relevant gain only with MB (recovering the performance gap with
respect to SpSh), an almost negligible gain with CCDM and HiDM, and
no gain with SpSh. Indeed, for small acceptance rate, the improvement
provided by sign-independent selection is no longer sufficient to
compensate for the information loss term in (\ref{eq:air}) \cite{secondini2022new},
so that all the dashed lines start to decrease. In practice, Fig.~\ref{fig:results_seqsel}(a)
shows that an effective sequence selection strategy requires the use
of a sign-dependent selection metric, while no relevant improvements
can be expected with sign-independent metrics (e.g., the EDI, LSAS,
NPN, and Kurtosis), unless the unbiased source (before selection)
is not properly optimized to remove intensity fluctuations. \footnote{For example, the EDI and LSAS do provide some nonlinear shaping gain
when used for a sort of sequence selection in \cite{askari2023probabilisticJLT}.
However the unbiased source employed in \cite{askari2023probabilisticJLT}
(PAS 256QAM and CCDM with rate $R_{\mathrm{DM}}=\unit[2.4]{bit/amplitude}$
and block length $904$) is not optimized for nonlinear performance,
and a similar gain could be obtained by using SpSh without sequence
selection. \emph{}} This is probably not a sufficient reason to use sequence selection
with a sign-independent metric, given that the same performance can
be achieved (with significantly lower complexity) by optimizing the
PAS blocklength and/or including CPR. On the other hand, Fig.~\ref{fig:results_seqsel}(b)
shows that sequence selection with unshaped but known signs (dotted-line
curves) achieves almost the same performance as sequence selection
with shaped signs. This means that shaping the signs is not really
necessary. They can remain uniform i.i.d. and it is sufficient to
account for them when selecting the amplitudes by using a sign-dependent
metric. In other words, for a given sequence of signs, there are good
and bad amplitude sequences that can be properly selected.

\begin{figure}
\centering \hfill{}(a)\hfill{}

\includegraphics[width=1\columnwidth]{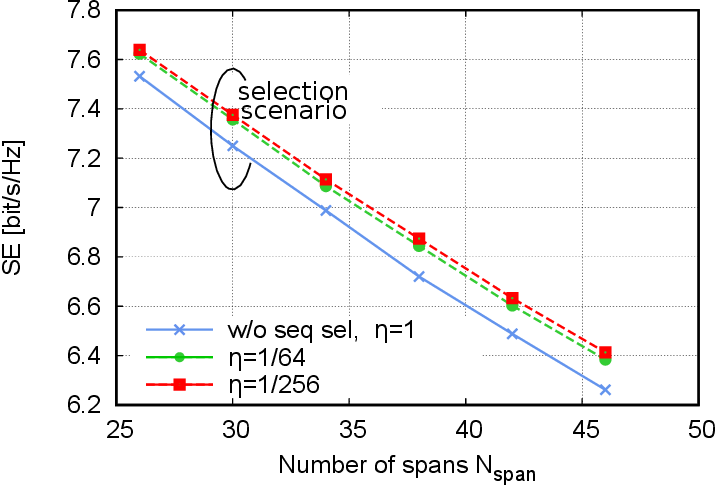}

\centering \hfill{}(b)\hfill{}

\includegraphics[width=1\columnwidth]{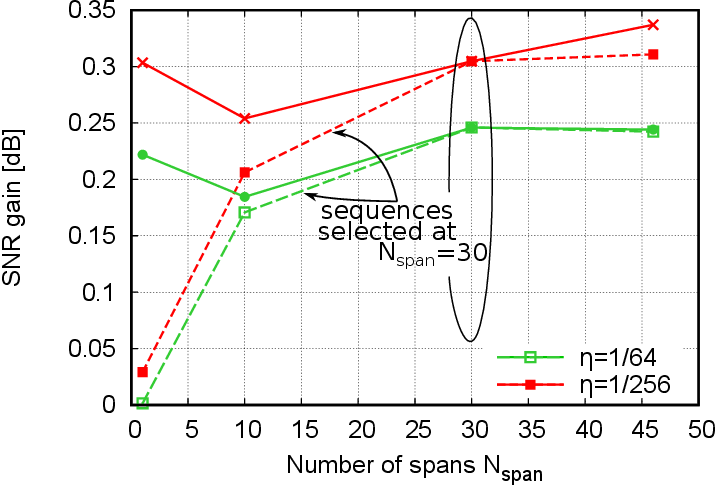}

\caption{\label{fig:results_seqsel_Nspan} Performance with sequence selection
(with shaped signs and SpSh as unbiased source) versus number of spans
$N_{\text{span}}$ when the sequences are selected for $N_{\text{span}}=30$
with dashed lines (a) SE, and (b) SNR gain, compared with performance
when sequences are selected for the actual $N_{\text{span}}$ with
solid lines.}
\end{figure}

\begin{figure}
\centering\includegraphics[width=1\columnwidth]{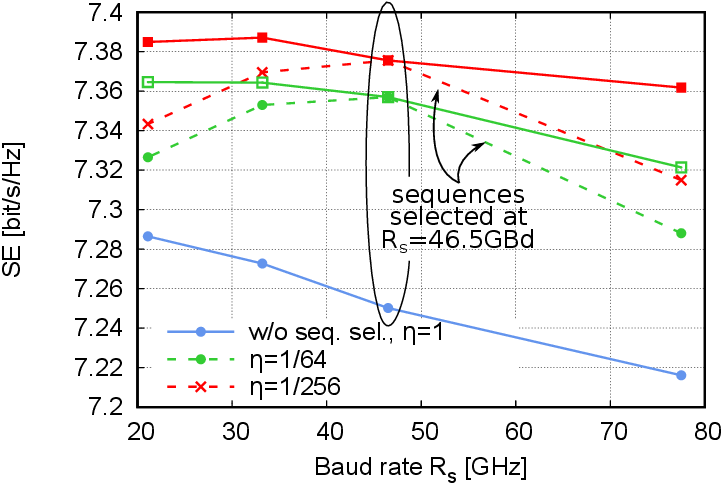}

\caption{\label{fig:results_seqsel_Rs} SE with sequence selection (with shaped
signs and SpSh as unbiased source) versus baud rate $R_{s}$ when
the sequences are selected for the actual rate (solid) or for $R_{s}=46.5$GBd
(dashed).}
\end{figure}

An obvious concern is how well the sequences selected for a given
scenario remain valid as the system parameters change. First of all,
the selection does not depend on the launch power, since the adopted
NLI selection metric simply scales, with good approximation, with
the cube of the power \cite{secondini2022new}, leaving unaltered
the sequence ranking (according to the NLI metric). Therefore, we
study the dependence of the selection on other system parameters,
namely the link length and the symbol rate, and using shaped signs
and SpSh at the unbiased source. Fig.~\ref{fig:results_seqsel_Nspan}(a)
shows the SE versus the number of spans $N_{\text{span}}$ at fixed
launch power $P=\unit[1]{dBm}$ when sequence selection is not applied
(cyan lines), and when sequence selection is applied with two different
acceptance rates. In this case, the sequence selection is done only
for $N_{\text{span}}=30$ and kept fixed for the other values of $N_{\text{span}}$.
Obviously, the performance of the system decreases when the link length
increases. Interestingly, the gain provided by sequence selection
remains nearly constant with $N_{\text{span}}$, despite the sequences
being selected for $30$ spans, meaning that the transmitter does
not need to know the exact link length and can use the same selection
criterion for different lengths. Obviously, when $N_{\text{span}}$
decreases, the SE saturates to the maximum value of $8.56$~bit/s/Hz,
and therefore we do not show the performance in this region in terms
of SE, since the gain would be limited by this saturation effect.
To study the behaviour for shorter links, we use a different performance
metric, the SNR, which does not saturate. Fig.~\ref{fig:results_seqsel_Nspan}(b)
shows the SNR gain (SNR with sequence selection minus SNR without
sequence selection), obtained with sequences selected at $N_{\text{span}}=30$
(the same in Fig.~\ref{fig:results_seqsel_Nspan}(a), shown with
dashed lines) and with sequences selected at the correct $N_{\text{span}}$
value (shown with solid lines). The figure shows that the gain decreases
and approaches zero when the sequences selected for $N_{\text{span}}=30$
are used for a much smaller link, while the gain achievable with the
sequences selected for the correct length remains almost constant
($\approx0.2$dB and $\approx0.3$dB with $\eta=1/64$ and $\eta=1/256$,
respectively).  On the other hand, Fig.\,\ref{fig:results_seqsel_Rs}
shows the SE versus the baud rate $R_{s}$. In this case, to keep
the overall spectral content unchanged and ensure a fair comparison,
we consider $d$ WDM channels with baud rate $R_{s}=232.5/d$\,GBd,
channel spacing $250/d$\,GHz, and launch power $P=3.32-\log_{2}d$\,dBm.
When the sequences are specifically selected for each given rate,
sequence selection provides similar gains for all the considered baud
rates, as shown with solid lines.\footnote{This is a preliminary result and we expect that with a proper optimization
of all the parameters (e.g., by increasing the sequence length and
decreasing the acceptance rate) sequence selection could provide a
larger gain at higher baud rates, being able to shape the signal on
a wider portion of the spectrum.} Conversely, when the sequences are selected for a specific rate (e.g.,
$R_{s}=\unit[46.5]{GBd}$ with $d=5$ channels) but used at different
rates, the gain has a maximum at this rate and reduces for the other
values, as shown with dashed lines. Still, the gain remains relevant
even if the baud rate changes significantly.


\subsection{Practical Implementation of Sequence Selection\label{subsec:Practical-implementation-seqsel}}

\begin{figure}
\includegraphics[width=1\columnwidth]{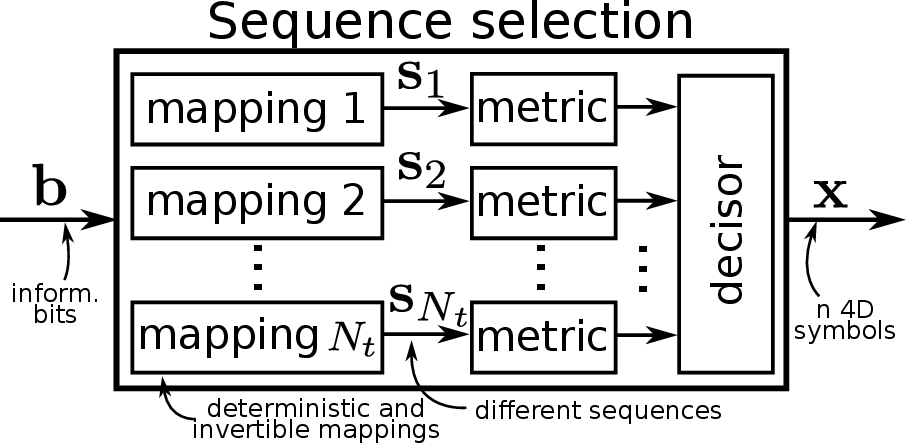}

\caption{\label{fig:pract_seq_sel} General structure for the practical implementation
of sequence selection.}
\end{figure}

The sequence selection technique described above serves as a nonlinear
equivalent of the MB distribution, allowing to establish the potential
and limitations of probabilistic shaping for nonlinear channels. Achieving
the AIR calculated with sequence selection requires a practical coded
modulation scheme to map information bits on sequence of symbols that
have the same statistical properties as those generated by the biased
source in Fig.~\ref{fig:seqsel}---exactly as the PAS scheme maps
information bits on (approximately) i.i.d. MB-distributed symbols
\cite{bocherer2015bandwidth}. A general scheme for a practical implementation
of sequence selection is sketched in Fig.\,\ref{fig:pract_seq_sel}.
A sequence of information bits $\mathbf{b}$ is mapped to $N_{t}$
different test sequences of symbols through $N_{t}$ different mapping
rules. The $N_{t}$ sequences are then compared according to a desired
metric, and the \emph{best} sequence $\mathbf{x}$ is selected for
transmission. A similar approach was previously proposed to reduce
the peak-to-average power ratio of orthogonal frequency division multiplexing
signals \cite{jayalath2000reducing}. The overall mapping strategy
should (i) be invertible, to allow decoding, and (ii) produce \emph{sufficiently}
different sequences, emulating the generation of $N_{t}$ independent
sequences. The main difference between ideal sequence selection in
Sec.\,\ref{subsec:Sequence-selection} and its practical implementation
is that the first selects sequences with rate $\eta$ according to
a certain threshold $\gamma_{\lambda}$ while the latter selects one
sequence out of $N_{t}$, without any constraints on its metric---and,
therefore, without control over its quality.

To the best of our knowledge, the first attempt to a practical implementation
of sequence selection for fiber nonlinearity mitigation is the list-encoding
CCDM structure \cite{wu2022list}. Using as a reference Fig.\,\ref{fig:pract_seq_sel},
the different sequences are generated by prepending some \emph{flipping}
bits to $\mathbf{b}$ and using the CCDM. $N_{t}$ different combinations
of $\log_{2}N_{t}$ flipping bits are used to obtain, at the CCDM
output, $N_{t}$ different sequences. The mapping strategy is easily
inverted by using the inverse CCDM and simply discarding the flipping
bits. In this case, the sequences are compared according to the EDI.
The list-encoding CCDM was later extended to account for different
metrics and different shaping techniques \cite{askari2022ecoc,askari2023probabilisticJLT}.
The main drawback of this scheme is that the generation of the sequences
is strictly related to the shaping strategy: a DM should be used,
the sequence length $n$ should be compatible with the DM block length,
and the selection takes into account only the amplitudes (and not
the signs).

However, this constraint can be overcome by separating the modulation
(which can include or not an initial shaping) and the generation of
different sequences, such as in the bit scrambling (BS) sequence selection
approach \cite{civelli2023practicalOFC}. In this case, given the
information bits $\mathbf{b}$, the $N_{t}$ sequences are generated
by (i) creating $N_{t}$ different scrambled versions of $\mathbf{b}$;
(ii) prepending a different combinations of $\log_{2}N_{t}$ pilot
bits to each of them (to allow decoding); and (iii) mapping the $N_{t}$
bit sequences to $N_{t}$ sequences of symbols according to a desired
(shaped or unshaped) modulation format. The BS scheme and other coded
modulation schemes with similar characteristics are described in the
next sections.

\subsubsection{Bit Scrambling Sequence Selection\label{subsec:BSSS}}

\begin{figure*}
\includegraphics[width=2\columnwidth]{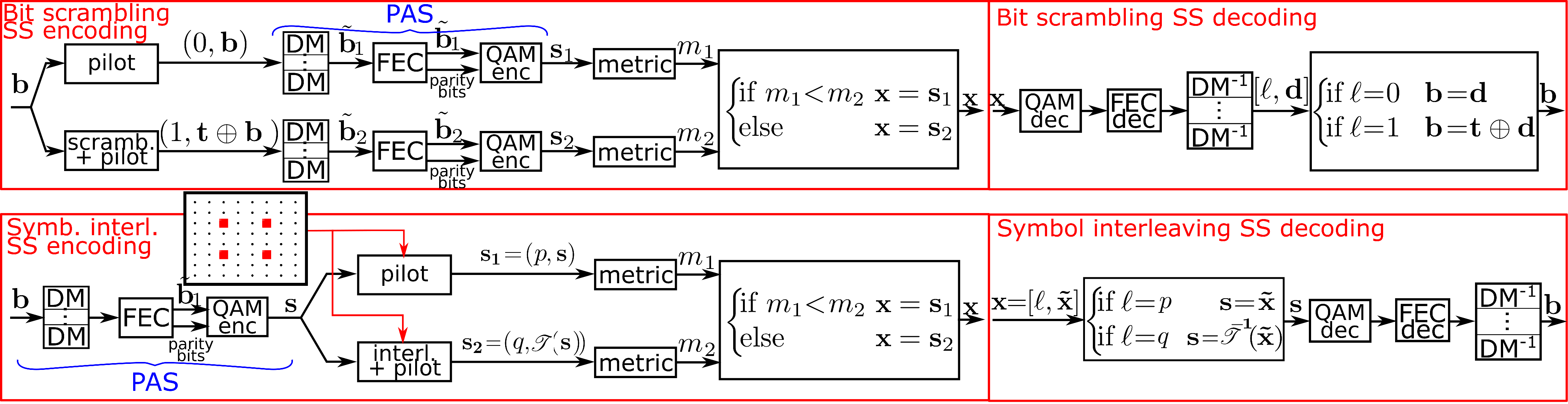}

\caption{\label{fig:BSSS_SISS} Bit scrambling (BS) and symbol interleaving
(SI) sequence selection}
\end{figure*}

The BS scheme is sketched on the top of Fig.\,\ref{fig:BSSS_SISS}
for $N_{t}=2$ test sequences, and works as follows \cite{civelli2023practicalOFC}.
Given a sequence of $n_{\text{inf}}$ information bits $\mathbf{b}$,
$N_{t}$ different sequences are generated, each obtained prepending
$n_{p}=\lceil\log_{2}N_{t}\rceil$ pilot bits to the sequence of bits
$\mathbf{t}_{k}\oplus\mathbf{b}$, with $\mathbf{t}_{k}$ being a
fixed sequence for $k=1,\dots,N_{t}$.\footnote{The optimization of the scrambling vectors $\mathbf{t}_{k}$ might
provide additional gain, but it is left for a future work.} For the sake of illustration, in the top of Fig.\,\ref{fig:BSSS_SISS}
the first sequence uses the pilot bit $0$ and $\mathbf{t}_{1}$ is
the identity, while the second uses the pilot bit $1$ and $\mathbf{t}_{2}=\mathbf{t}$.
Each bit sequence, of length $n_{\text{inf}}+n_{p}$, is sent to a
conventional PAS block to produce $n$ 4D QAM symbols. The PAS block
is the concatenation of DM, FEC with rate $c$, and QAM encoder, with
the following details. Firstly, the sequence of bits is sent to a
set of $4n/N_{\text{DM}}$ DMs, each with block length $N_{\text{DM}}$
and rate $R_{\text{DM}}$. The set of DMs takes $4nR_{\text{DM}}$
bits to produce $4n$ shaped amplitudes from an alphabet of $\sqrt{M}/2$
elements (corresponding to $4n\log_{2}(\sqrt{M}/2)$ bits), while
the remaining $n_{u}=n_{\text{inf}}+n_{p}-4nR_{\text{DM}}$ bits are
left unchanged. The resulting sequences, of $n_{\tilde{b}}=n_{u}+4n\log_{2}(\sqrt{M}/2)$
bits---indicated as $\tilde{\mathbf{b}}_{1}$ and $\tilde{\mathbf{b}}_{2}$
in Fig.\,\ref{fig:BSSS_SISS}---, are sent to the FEC encoder.
The FEC produces $n_{\tilde{b}}(1-c)/c$ parity bits, which, together
with the input $n_{\tilde{b}}$ bits, generate a sequence of $n$
4D QAM symbols according to PAS: $4n$ shaped amplitudes (written
as $4n\log_{2}(\sqrt{M}/2)$ bits), and $4n$ signs coming from $n_{\tilde{b}}(1-c)/c$
parity bits and from the unshaped $n_{u}$ bits. After some easy calculation
it turns out that $n_{u}=2n\left(2-\left(1-c\right)\log_{2}M\right)$,
$n_{\tilde{b}}=2nc\log_{2}M$, and $n_{\text{inf}}=2n\left(R_{4\text{D}}/2-(1-c)\log_{2}M\right)-n_{p}$.\footnote{Note that, indeed, $2n\left(R_{4\text{D}}/2-(1-c)\log_{2}M\right)$
is exactly the amount of bits that are encoded on $n$ 4D symbols
with PAS, when sequence selection is not used.} Finally, the metric corresponding to each sequence of $n$ 4D QAM
symbols is evaluated, and the sequence with the best metric is transmitted.
At the receiver side, QAM demodulation, FEC decoding, and inverse
DM are performed as in the standard PAS scheme, and the pilot bits
are finally used to descramble the received bits and recover $\mathbf{b}$.

The main difference with respect to list-encoding CCDM and its variants
is that the generation of different sequences is not induced by the
DM (hence, combined with PAS and constrained by it), but simply obtained
with bit scramblers (xor operations). As a consequence, the length
$n$ of the sequences to be selected is not related to the PAS block
length (except for the soft requirement that $4n/N_{\text{DM}}$ should
be an integer number), and the two lengths can be optimized almost
independently. Moreover, the selection process can involve also the
sign bits of the sequence, which are not involved in PAS, allowing
for higher gains (as shown in Fig.~\ref{fig:results_seqsel}). 

The main drawback of the BS sequence selection is its complexity,
since FEC, DM and metric should be evaluated for each $N_{t}$ sequence.
In particular, the FEC should be performed for each sequence since
the parity bits are sent to the signs and depends on the amplitudes.
This implies that the number of bits in output from the DMs block,
equal to $2nc\log_{2}M$, should be compatible with the input length
to the FEC. For example, the low-density parity-check (LDPC) codes
from DVB-S.2 standard with rate $1/2$ takes in input $32400$ bits,
and, therefore, requires sequences of length $n=5400$ if $M=64$.

\subsubsection{Symbol Interleaving Sequence Selection}

The SI approach, sketched on the bottom of Fig.\,\ref{fig:BSSS_SISS}
for $N_{t}=2$ test sequences, was proposed to overcome some of the
drawbacks of BS \cite{civelli2023practicalOFC}. The main difference
with respect to BS is that the selection process is moved after the
cascade of DM, FEC, and QAM encoder---the PAS block. The $N_{t}$
test sequences are generated by applying different interleavers to
the modulated sequence, and adding different pilot symbols to each
test sequence---selected from a subset of the employed QAM constellation---to
identify the interleaving operation and allow deinterleaving at the
receiver. Therefore, the subset of QAM symbols should be large enough
to identify the $N_{t}$ different sequences by using a small number
of pilot symbols, but also have a large minimum distance between the
symbols to minimize the error probability (given that the pilot symbols
are not protected by FEC). In our case, we considered a set of $4$
symbols on both polarizations, as shown in the inset in the bottom
of Fig.\,\ref{fig:BSSS_SISS}, allowing to address up to $16$ different
sequences with one 4D symbol, and up to 256 with two 4D symbols. With
this choice we did not found any error on the pilot symbols during
our simulations.

Overall, this approach allows to reduce the complexity by applying
the DM and FEC only once, rather than $N_{t}$ times, and to avoid
any constraint on the FEC and sequence selection lengths. However,
the pilot symbols are not protected by FEC, and their use induces
a higher rate loss compared to the $\lceil\log_{2}N_{t}\rceil$ bits
per sequence required by BS. For example, choosing a subset of $4$
symbols out of a $64$QAM constellation as in our case, one needs
$\lceil(\log_{2}N_{t})/4\rceil$ $4$D pilot symbols, with a loss
of $12\lceil(\log_{2}N_{t})/4\rceil$ bits per sequence.\footnote{Furthermore, intearleaving $4n/N_{\text{DM}}\geq1$ output DM blocks
reduces the nonlinear shaping gain due to short block length DM, thus
reducing the overall performance. This requires to use smarter interleaving
approaches, for example intearleaving the sequence according to independent
blocks, each of length $N_{\text{DM}}$.}

\subsubsection{FEC-Independent BS\label{subsec:FEC-Independent-BS}}

\begin{figure*}
\centering\includegraphics[width=2\columnwidth]{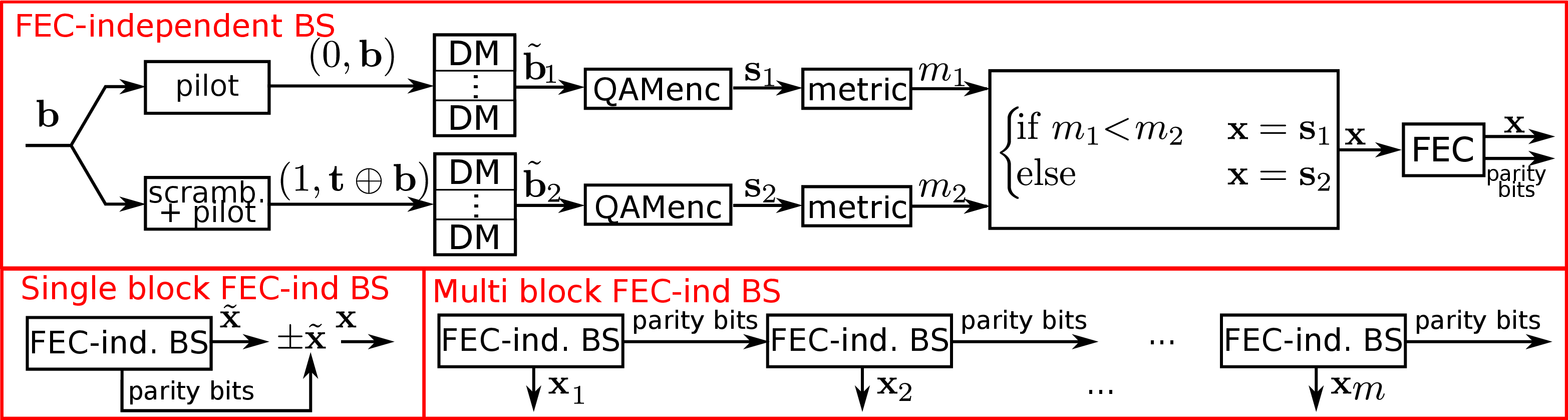}

\caption{\label{fig:schemeFECind} FEC-independent BS sequence selection (top)
with single block (SB) (bottom left) and multi block (MB) (bottom
right) implementations.}
\end{figure*}

A general approach to solve the issues related to the interaction
of FEC in the BS approach is the FEC-independent BS, sketched in Fig.\,\ref{fig:schemeFECind}.
In a nutshell, the FEC-independent BS uses the BS approach (hence
without the issues of the SI approach), but moving the FEC \emph{after}
the selection process. In this way, the FEC is applied only to the
selected sequence (rather than to $N_{t}$ sequences). Furthermore,
the lengths of DM, sequences, and FEC can be independently selected,
since the FEC can be applied on the concatenation of several sequences.
In this scheme, the main issue is that of finding a proper way to
map the parity bits generated by the FEC on the selected sequence,
without altering the particular shaping induced by the selection process.
The problem is not simple, since the number of parity bits can be
quite large. Given a sequence of $n$ 4D symbols, an $M$-QAM constellation,
and a FEC rate $c$, the number of parity bits is $2n\log_{2}M(1-c)$,
and the ratio between the parity bits and the signs bits is $\nu=(1-c)\log_{2}M/2$.
For example, with $M=64$: if $c=2/3$, $\nu=100$\%, while if $c=4/5$,
$\nu=60\%$. Possible solutions for the parity bits are discussed
below.

The simplest approach is to use a sign-independent metric for the
selection and place the parity bits on the signs of the symbols (as
in conventional PAS). In this case, the BS and the FEC-independent
BS approach are equivalent, since the metric value is not altered
by the specific parity bits produced by the FEC, so that they can
be determined and mapped to the sequence after the selection process.
Unfortunately, this approach is not practically useful, since it does
not provide any additional gain with respect to a properly optimized
conventional PAS, as shown by the case of ideal sequence selection
with unshaped unknown signs in Fig.~\ref{fig:results_seqsel}.

A possible approach to use a sign-dependent metric is the single-block
FEC-independent BS (SB-BS) scheme sketched in the lower left of Fig.\,\ref{fig:schemeFECind}.
In this case, the sequence of symbols (including their signs) is selected
according to the desired sign-dependent metric, but then the parity
bits produced by the FEC encoder are mapped to (part of) the signs
of the same selected sequence, actually modifying them with respect
to the selected ones, hence reducing the accuracy of the selection
process. When the FEC rate is small, the percentage of parity bits
with respect to the sign bits $\nu$ is large (e.g., $c=2/3$, $\nu=100$\%)
and the method practically reduces to the use of a sign-independent
metric (with no expected additional gain with respect to conventional
PAS). However, when the FEC rate is larger and $\nu$ becomes smaller,
part of the sequence is correctly characterized by the metric and
the method should provide some gain over the use of a sign-independent
metric. The position of the parity bits is discussed in the next Section.

An alternative approach to FEC-independent BS is suggested by the
scenario of ideal sequence selection with \emph{unshaped} and \emph{known}
signs considered in Section~\ref{subsec:Sequence-selection}, which
shows that it is not necessary to select (shape) the signs of the
sequence, since it is sufficient to know their value when selecting
the amplitudes to achieve nearly the same shaping gain. A possible
implementation of such a strategy is the multi-block FEC-independent
BS (MB-BS) sketched in the lower right of Fig.\,\ref{fig:schemeFECind},
where the parity bits of each selected sequence are mapped to a fraction
$\nu$ of the signs of the next sequence (similarly to the \emph{bootstrap}
scheme proposed in \cite{bocherer2011operating}). In this case,
when selecting a certain sequence, these signs are predetermined by
the parity bits coming from the previous sequence; they are known
and can be used in the computation of the metric, but must be kept
fixed during the selection process (the corresponding bits are not
scrambled). The main drawback of MB-BS is the latency, since the encoding
encompasses several blocks, and the decoding should be performed in
reverse order, after the last block has been received.

\subsubsection{Performance of Sequence Selection Implementations}

\begin{figure}
\centering \includegraphics[width=1\columnwidth]{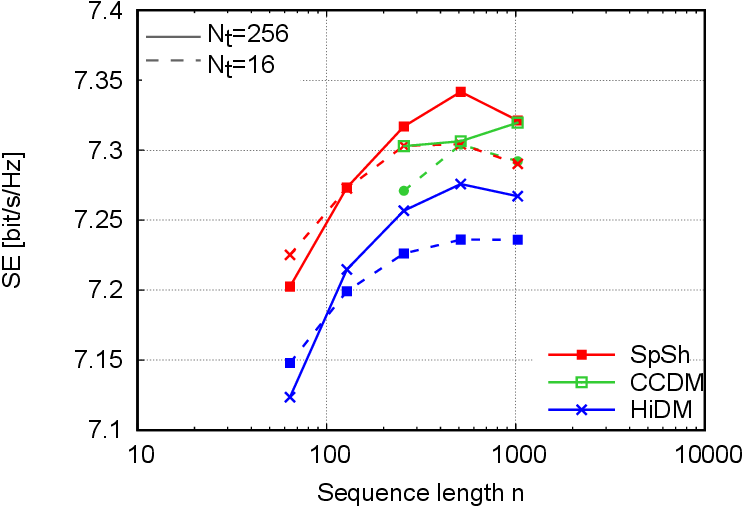}

\caption{\label{fig:results1}Performance of BS versus $n$.}
\end{figure}

In this Section, the performance of different sequence selection schemes
is studied. At the unbiased source, we consider the same four different
shaping techniques as in Sec.\,\ref{subsec:Sequence-selection}.
As a metric, we consider the block-memoryless single-channel NLI metric
in \cite[eq. (13)]{secondini2022new}, computed according to the algorithm
described in \cite[Sec. V-B]{secondini2022new} (but extended to two
polarizations). This metric is simpler and more practical than the
one adopted in the ideal case, since it requires the propagation of
a shorter sequence and avoids averaging over different realizations.
However, it neglects inter-block NLI, so that it is expected to provide
a less accurate selection and, consequently, a lower gain. To assess
the performance, we evaluate the SE at the optimal power $P=\unit[1]{dBm}$,
through the AIR in Eq.\,(\ref{eq:air}), recalling that $N_{p}/N_{a}=N_{t}$
when practical implementations of sequence selection are considered.
To partially compensate for the loss due to selection, for MB, SpSh,
and CCDM, we increase the PAS rate by the same quantity $\frac{1}{n}\log_{2}(N_{p}/N_{a})$
in (\ref{eq:air}). Regarding the HiDM, for the sake of simplicity,
we use the HiDM structure in Sec.\,\ref{sec:Probabilistic-constellation-shap}
for $N_{t}\leq16$, while for $N_{t}>16$, we increase the number
of input bits to each LUT of each layer to $(7,3,5,4,3,9)$, to obtain
a larger rate equal to $1.3398$\,bit/amplitude (with negligible
difference in rate loss and memory requirements).

Fig.\,\ref{fig:results1} shows the performance of BS sequence selection---here
and in the following, we use BS to refer to the simple (not FEC-independent)
implementation described in Section\,\ref{subsec:BSSS}, and SB-BS
and MB-BS for the FEC-independent implementations described in Section\,\ref{subsec:FEC-Independent-BS}---as
a function of the sequence length $n$\footnote{For the CCDM, we consider only $n\geq256$, since $N_{\text{DM}}=1024$.}
for $N_{t}=256$ (solid) and $N_{t}=16$ (dashed) tested sequences.
When $n$ is too small, sequence selection does not work because the
loss induced by selection ($\frac{1}{n}\log_{2}N_{t}$) is large and
the impact of adjacent sequences (not considered by the metric) too
relevant. On the other hand, when $n$ is increased, sequence selection
yields a larger gain, provided that a sufficiently small acceptance
rate is considered \cite{secondini2022new}. The figure shows that
a good trade-off between these two effects is obtained with $n=512$,
which will be considered in the following. Note that in the list-encoding
CCDM approach, $n$ is strictly related to $N_{\text{DM}}$ and this
optimization is not possible.

\begin{figure}
\centering \hfill{}(a)\hfill{}

\includegraphics[width=1\columnwidth]{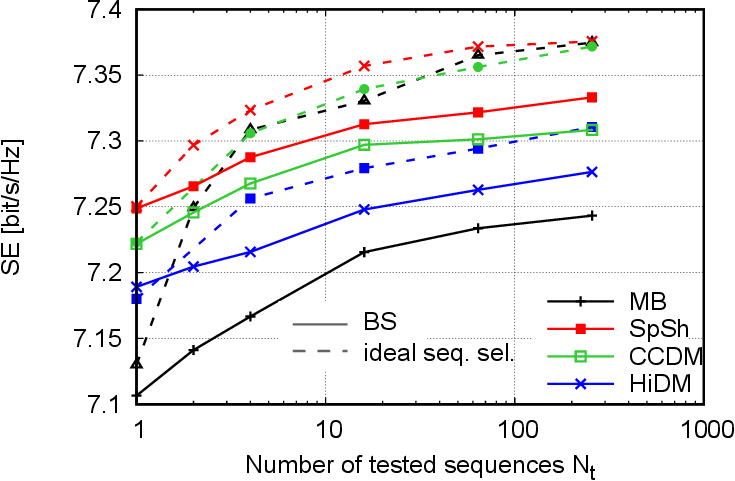}

\centering \hfill{}(b)\hfill{}

\includegraphics[width=1\columnwidth]{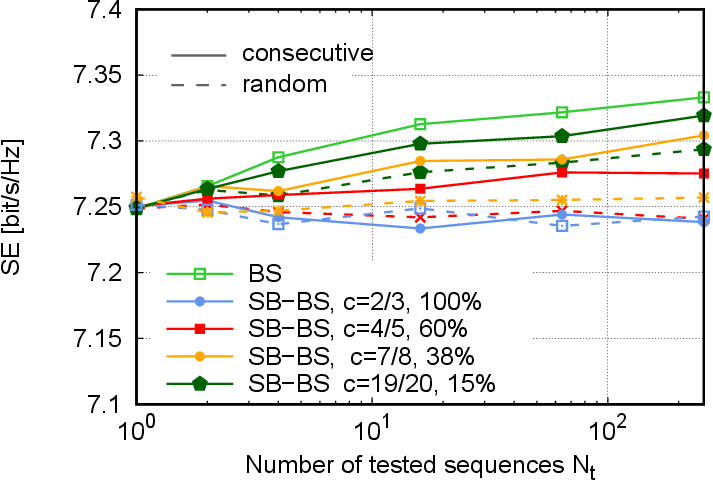}

\centering \hfill{}(c)\hfill{}

\includegraphics[width=1\columnwidth]{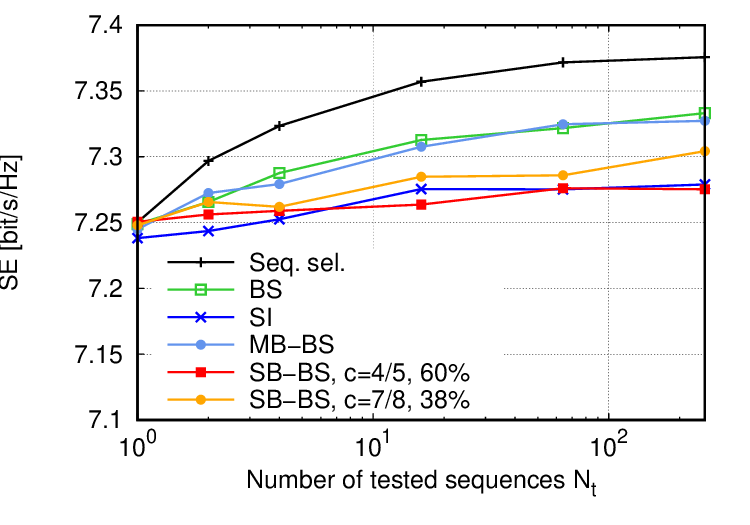}

\caption{\label{fig:results2}Performance of different sequence selection techniques
for a fixed blocklength $n=512$: (a) BS (solid) and ideal (dashed)
sequence selection combined with different PAS schemes; (b) SB-BS
with SpSh-based PAS and different allocation of parity bits (solid:
consecutive, dashed: random); (c) all techniques compared (with SpSh-based
PAS).}
\end{figure}

Figure\,\ref{fig:results2}(a) compares the performance of BS sequence
selection with that of ideal sequence selection with shaped signs
(already shown in Fig.~\ref{fig:results_seqsel}), as a function
of $N_{t}$ and for different unbiased sources. While the performance
improves with $N_{t}$ in all cases, the difference between BS and
ideal sequence selection is not negligible. We believe that this difference
is mainly due to the different metric employed in the two cases---more
accurate in the ideal case, simpler in the BS case---and that more
efficient practical implementations able to close this gap are possible.
The BS implementation yields a similar SE gain (up to approximately
0.09~bit/s/Hz) with any of the three DMs (SpSh, CCDM, and HiDM),
and a larger gain (about 0.14~bit/s/Hz) with i.i.d. MB symbols, for
which it is also able to partly recover the performance gap with respect
to the finite-blocklength DMs (as already observed in the ideal case
but to a lesser extent).

Fig.\,\ref{fig:results2}(b) shows the performance obtained with
the SB-BS sequence selection for different parity-to-sign-bit ratios
$\nu$ (corresponding to different FEC rates $c$) and two different
allocation schemes of the parity bits: \emph{consecutive}, when the
parity bits are all mapped to the signs (for both quadratures and
polarizations) of the first consecutive symbols of the sequence; and
\emph{random}, when they are randomly allocated along the sequence
on a fraction $\nu$ of the signs. For the sake of comparison, the
performance of BS is also reported. As expected, the figure shows
that the performance improves as the percentage of parity bit decreases
(and the FEC rate consequently increases), and there is no gain when
$\nu=100$\%. Interestingly, placing the parity bits on consecutive
symbols provides better performance, meaning that it is better to
have a part of the sequence (the one without parity) properly shaped
and another part not (the one modified by the parity bits), than having
a uniform distribution of modified signs over the whole sequence.
In the following, we will consider only the case with consecutive
symbols for the SB-BS approach.

Next, Fig.\,\ref{fig:results2}(c) compares the performance of all
the sequence selection approaches discussed so far, including MB-BS
and SI, considering the SpSh DM in all the cases. Interestingly, the
FEC-independent scheme based on the MB-BS approach is able to achieve
the same performance as the original BS, independently of the fraction
$\nu$ of sign bits that are predetermined by the parity bits of the
previous block and of their allocation scheme---which are, therefore,
not indicated in the legend. This result is in accordance with the
behavior shown for ideal sequence selection with unshaped known signs
in Section\,\ref{subsec:Sequence-selection} and implies that FEC
can actually be moved after sequence selection (hence reducing the
complexity) with no performance penalty, provided that a longer latency
can be tolerated. On the other hand, both SB-BS and SI yield a lower
SE gain, the former performing approximately as the latter for $c=4/5$.

\begin{figure}
\centering \includegraphics[width=1\columnwidth]{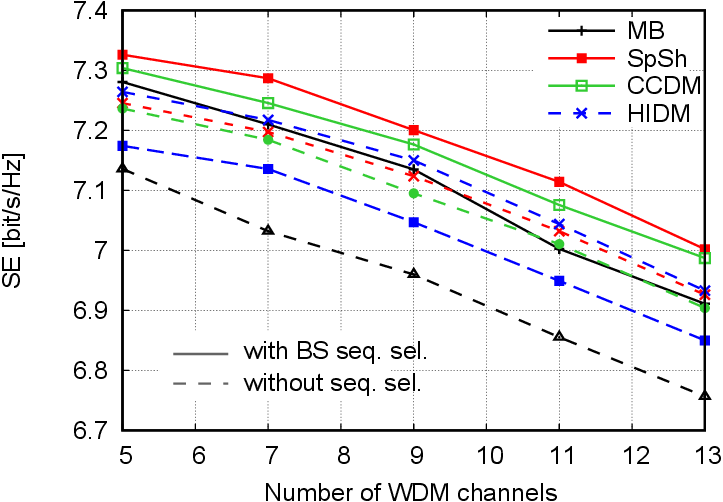}

\caption{\label{fig:results1_new}Performance with and without BS sequence
selection versus number of WDM channels.}
\end{figure}

Finally, to check the dependency of the gain due to sequence selection
on the WDM signal bandwidth, which affects the nonlinear penalty,
Fig.\,\ref{fig:results1_new} shows the SE as a function of the number
of WDM channels, when using BS. The performance is shown at $P=\unit[1]{dBm},$which
remains approximately the optimal power for the case without sequence
selection. As expected, the figure shows that when the number of WDM
channels increases, the performance of all methods worsens, as a consequence
of the increased nonlinearity. However, the gain provided by sequence
selection over the corresponding unbiased source remains nearly constant
(at least up to the case of 13 WDM channels considered in our simulations).

\subsection{Interaction of Sequence Selection and Carrier Phase Recovery}

The study of the interaction of short-block-length PAS and carrier
phase recovery (CPR) in the nonlinear regime showed that, in the presence
of the latter, the nonlinear shaping gain provided by the former vanishes
or becomes negligible in most of the scenarios of interest \cite{civelli2020interplayECOC,civelli2023JLTNPN}.
In fact, CPR, always present in practical coherent systems, can already
mitigate some nonlinear effects, possibly making the work done by
a specific nonlinear constellation shaping algorithm less effective
or even useless. This result pushed the research toward new shaping
architectures specifically suited for nonlinearity mitigation, such
as sequence selection, and made clear the need to include the effect
of CPR in the assessment of their performance.

\begin{figure}
\includegraphics[width=1\columnwidth]{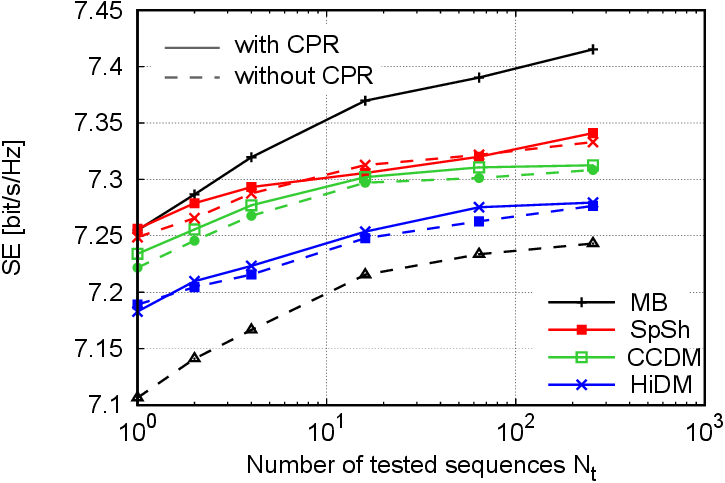}

\caption{\label{fig:resultsBPS}Impact of CPR on the performance of BS combined
with different PAS schemes.}
\end{figure}

In this Section, we add a CPR at the receiver side, implemented with
the blind phase search (BPS) algorithm with $64$ test angles and
with a window of $281$ symbols ($140$ on each side), previously
optimized to obtain the best performance \cite{pfau2009BPS,civelli2023JLTNPN}.\footnote{For the sake of simplicity, we do not include laser phase noise in
the system. However, we expect the overall result not to change, as
in \cite{civelli2023JLTNPN}.} Figure\,\ref{fig:resultsBPS} shows the peak SE (at optimal launch
power $P=\unit[1]{dBm}$) versus number of tested sequences when the
CPR is included (solid) or not (dashed), with the BS approach and
$n=512$. The baseline performance without sequence selection improves
significantly for i.i.d. MB-distributed symbols (whose intensity fluctuations
cause a relevant nonlinear phase noise that can be mitigated by the
CPR), whereas it does not improve or has a small improvement when
a short-block-length DM is used (which already produces signals with
less intensity fluctuations)---for more details about this refer
to \cite{civelli2023JLTNPN}. Conversely, the figure shows that sequence
selection is able to provide approximately the same gain when CPR
is included or not, in all cases. When sequence selection is not applied
($N_{t}=1$), the difference between MB and the various DMs is due
to both the rate loss and the different characteristics in terms of
NLI generation. For instance, for $N_{t}=1$, SpSh and MB have nearly
the same performance, meaning that the rate loss of SpSh is compensated
by its advantage (small, when CPR is included, but non-negligible)
over MB in terms of NLI generation. However, when the acceptance rate
decreases, sequence selection discards the sequences with high NLI,
so that the NLI advantage of SpSh (and of the other DMs) over MB vanishes,
and the only remaining effect is the rate loss, which (once reported
in bit/s/Hz) explains the SE differences between MB and practical
DMs. Overall, this important result implies that the nonlinear shaping
gain provided by sequence selection remains when CPR is included,
differently from the nonlinear shaping gain obtained with short block
length DM. 

\section{Discussion and Conclusion\label{sec:conclusion}}

In this work the concept of probabilistic constellation shaping for
nonlinear channels was discussed. In a nutshell, probabilistic shaping
aims at improving the performance of a fiber optic communication system
by encoding information onto the symbols of a QAM constellation in
a smart way, taking into account the interaction of adjacent symbols
in all possible dimensions (time, quadratures and polarizations).
Sequence selection was used to assess the potential of probabilistic
shaping for nonlinear channels and address its implementation.

Sequence selection can be seen as a\emph{ generalization} of conventional
probabilistic shaping in the absence of a priori knowledge about (i)
optimal distribution, (ii) features or characteristics of a good shaping,
and (iii) optimal bit-to-symbol mapping rules. Basically, sequence
selection produces properly shaped sequences of symbols, starting
from symbols generated by any source, referred to as \emph{unbiased}
source, and using a proper performance metric. Using ideal sequence
selection as an upper bound, we have drawn some important conclusions
about practical implementations of probabilistic shaping for nonlinear
channels: (i) if the unbiased source is already optimized to minimize
the intensity fluctuations of the signal (such as with short-block-length
PAS), a further optimization of the transmitted sequences based on
a sign-independent metric does not provide any additional improvement;
(ii) full shaping of both amplitudes and signs achieves the best performance
(with a gain of about $0.13$\,bits/s/Hz in a typical scenario);
(iii) nearly the same performance can be achieved by shaping only
the amplitudes, provided that the amplitudes are shaped depending
on the values taken by the signs (i.e., by using a sign-dependent
metric for sequence selection). Furthermore, we have shown that the
optimal shaping depends only mildly on system parameters such as baud
rate and number of spans. In fact, a lower but still relevant gain
can be obtained if those parameters are changed (within a reasonable
range) without reoptimizing the shaping, allowing for a good amount
of flexibility in the system optimization. Moreover, we have shown
that the gain provided by BS sequence selection remains approximately
constant when the number of WDM channels increases.

Different techniques for the practical implementation of sequence
selection have been discussed, highlighting their advantages and disadvantages.
The bit scrambling (BS) sequence selection allows to obtain good nonlinear
shaping gain, but has some issues related to the interaction with
FEC. These issues can be overcome with the symbol interleaving (SI)
approach, which, however, significantly hampers its performance. Therefore,
we proposed two variations of BS for a FEC-independent approach: the
single-block BS (SB-BS) and the multi-block BS (MB-BS). On the one
hand, the SB-BS is a straightforward variation of BS , which entails
only a limited performance degradation compared to BS when combined
with a high-rate FEC; on the other hand, the MB-BS approach concatenates
several BS blocks and is able to achieve the same performance as BS,
regardless of the FEC rate, at the expense of increased latency.

The main issue affecting all sequence selection implementations and
preventing their actual implementation is the complexity of the metric.
Indeed, the metric should be evaluated for each test sequence, and
the number of test sequences $N_{t}$ should be large enough to ensure
sufficiently good performance. Unfortunately, currently available
low-complexity metrics---e.g., the EDI, LSAS, NPN, and windowed Kurtosis---do
not account for the signs of the symbols, so that a sequence selection
based on one of those metrics would not grant any additional gain
compared to a properly optimized conventional PAS. In this manuscript,
we considered a complex metric based on the split-step Fourier method
to accurately evaluate the nonlinear interference generated by each
sequence. This approach, though too complex for a practical implementation,
allowed us to perform this study and draw several interesting conclusions
and guidelines about the implementation of sequence selection. However,
the adopted metric provides an accurate estimation of the different
nonlinear interference affecting each symbol of the sequence, which
is a much more detailed knowledge than the one actually needed for
an efficient selection. In fact, a simpler metric able to \emph{rank}
the sequences according to their \emph{average} nonlinear interference
would be sufficient to obtain the same performance. The design of
such a low-complexity metric is left for a future work, together with
the investigation about metrics that take into account the nonlinear
interference caused by adjacent WDM channels and signal-noise interactions
due to EDFA.

Finally, we have evaluated the performance of sequence selection when
carrier phase recovery (CPR) is included in the system. Indeed, CPR
is always present in practical systems and can already compensate
for part of the nonlinearity, often reducing (or even eliminating)
the additional improvement achievable with other mitigation methods,
such as short-block-length PAS. Conversely, we showed in this manuscript
that the nonlinear shaping gain provided by sequence selection remains
almost unchanged when CPR is included in the system. This result encourages
the use of sequence selection for nonlinear constellation shaping.

\section*{Acknowledgment}

This work was partially supported by the European Union under the
Italian National Recovery and Resilience Plan (NRRP) of NextGenerationEU,
partnership on \textquotedblleft Telecommunications of the Future\textquotedblright{}
(PE00000001 - program \textquotedblleft RESTART\textquotedblright ).

\bibliographystyle{ieeetr}

\end{document}